\documentclass[
aps,
pra,
showpacs
amsmath,
amssymb,
preprint,
floatfix,
lengthcheck, %two columns
superscriptaddress
]{revtex4-1}

\usepackage{graphicx}
\usepackage{mathtools}
\usepackage{braket}
\usepackage{hyperref}
\usepackage{bbold}
\usepackage{calrsfs}
\usepackage{dsfont}
\usepackage{cancel}
\usepackage{mathrsfs}
\usepackage[margin=0.75in]{geometry}
\usepackage[normalem]{ulem}
\usepackage{multirow}

\usepackage{color}

\DeclareMathAlphabet{\pazocal}{OMS}{zplm}{m}{n}

\makeatletter \renewcommand\@make@capt@title[2]{%
\@ifx@empty\float@link{\@firstofone}{\expandafter\href\expandafter{\float@link}}%
\sffamily{\textbf{#1}}\@caption@fignum@sep#2 }% \makeatother
\usepackage{placeins} % Allows use of the command \FloatBarrier to prevent figures and tables from moving past that point

\definecolor{nrppurple}{RGB}{128,0,128}
\usepackage{hyperref}% add hypertext capabilities
\hypersetup{
colorlinks=true,
	linkcolor=nrppurple,
	citecolor=nrppurple,
	urlcolor=nrppurple,
}

\begin{document}

%-----------------------------------------------------------------
%                      Title and Authors                       
%-----------------------------------------------------------------

\title{Nonlinear optical processes in centrosymmetric systems by \\ strong-coupling-induced symmetry breaking}

\author{Davis M. Welakuh}
\email[Electronic address:\;]{dwelakuh@seas.harvard.edu}
\affiliation{Harvard John A. Paulson School Of Engineering And Applied Sciences, Harvard University, Cambridge, Massachusetts 02138, USA}

\author{Prineha Narang}
\email[Electronic address:\;]{prineha@seas.harvard.edu}
\affiliation{Harvard John A. Paulson School Of Engineering And Applied Sciences, Harvard University, Cambridge, Massachusetts 02138, USA}

%-----------------------------------------------------------------
%                      Abstract                       
%-----------------------------------------------------------------

\begin{abstract}
Nonlinear optical processes associated with even-order nonlinear susceptibilities are critical for both classical and quantum technologies. Inversion symmetry, however, prevents nonlinear optical responses mediated by even-order susceptibilities in several material systems pertinent for applications in nanophotonics. Here, we demonstrate induced nonlinear optical processes, namely second- and fourth-harmonic generation that are naturally forbidden in an inversion symmetric system, by strongly coupling to a photon mode of a high-Q optical cavity. As an illustrative system with an inversion symmetry, we consider a semiconductor quantum ring of GaAs that features a single effective electron. For the coupled system, we control the inversion symmetry breaking by changing the light-matter coupling strength which at the same time allows to tune the nonlinear conversion efficiency. We find that the harmonic generation yield can be significantly increased by increasing the light-matter coupling strength in an experimentally feasible way. In the few-photon limit where the incident pump field is a coherent state with just a few photons, we find that the harmonic conversion efficiency is increased for strong coupling as opposed to using intense pump fields. This new approach is applicable to a wide variety of centrosymmetric systems as the symmetry breaking rest on the properties of the photonic environment used to achieve strong light-matter interaction. Our work constitutes a step forward in the direction of realizing physically forbidden nonlinear optical processes in centrosymmetric materials widely adopted for applications in integrated photonics.
\end{abstract}

\maketitle

%\tableofcontents

%\section{Introduction}
%\label{sec:introduction}

The interaction between light and matter is highly dependent on the symmetry of the physical system, across atoms, molecules, and solids, which dictates the linear and nonlinear optical response. Material systems that possess an inversion symmetry center (that is, centrosymmetric systems) have a negligible nonlinear response for processes associated with even-order susceptibilities $\chi^{(n)}$ (where $n$ are even natural numbers) as these quantities vanish~\cite{boyd1992} by symmetry. Nonetheless, nonlinear optical processes associated with even-order $\chi^{(n)}$ in centrosymmetric systems can still be observed by breaking the symmetry. There exist several ways to achieve this symmetry breaking, for example, by applying external strain~\cite{jacobsen2006,cazzanelli2012} or from twisted interfaces in van der Waals heterostructures~\cite{yao2021}, to using DC electric fields~\cite{timurdogan2017,klein2017}, or using strong THz pulses~\cite{grishunin2017,vampa2018}, or even confining in optical~\cite{breunig2016,gouveia2013,lettieri2005,zhang2019a} or a plasmonic cavities~\cite{li2021}. Importantly, these different approaches of symmetry breaking require high or relatively high~\cite{zhang2019a} pump powers for efficient harmonic generation.

Strong light-matter interaction within confined electromagnetic environments such as optical cavities or plasmonic devices provide a parallel way to control or modify material properties~\cite{flick2018a,ruggenthaler2017b}. In the strong coupling regime where energy is coherently exchanged between light and matter, a mode hybridizes with a material resonance leading to the formation of hybrid light-matter states known as polaritons. This regime has been explored to control chemical reactions~\cite{hutchison2012}, realize polariton lasing~\cite{cohen2010} and even the enhancement of harmonic generation from polaritonic states in noncentrosymmetric systems~\cite{chervy2016,barachati2018,welakuh2022b}. Such control mediated by strong light-matter coupling is suitable to investigate nonlinear optical processes associated with even-order susceptibilities in centrosymmetric systems with the possibility for control of the efficiency.

In this \emph{Letter}, we present a general approach to break the inversion symmetry of material systems which allows for highly tunable even-order harmonic generation naturally forbidden in such systems. Our approach relies on a quantized treatment of the coupled light-matter system where the incident (pump) field is treated as a quantized field in a coherent state. When the light-matter system is strongly coupled, this leads to an induced symmetry-breaking, thus, allowing for nonlinear optical processes associated with even-order susceptibilities. An important feature of our mechanism is that we can readily control the inversion symmetry-breaking. Such control is achieved by increasing the coupling strength in an experimentally feasible way which increases the degree of the induced asymmetry and simultaneously result in an efficient ever-order harmonic yield from the strongly coupled light-matter system. Several applications ranging from optical frequency conversion to imaging and biosensing where control over efficiency is important~\cite{garmire2013} could leverage this mechanism. An advantage of our approach is that we work in the few-photon limit realized in quantum optical experiments~\cite{chang2014,hacker2019} which requires weak pump power as opposed to methods that require high pump power~\cite{breunig2016,gouveia2013,lettieri2005}. With this we show that the nonlinear conversion efficiency increases using the few-photon limit with strong coupling as opposed to using intense pump fields. An added merit of the quantized treatment is that we can investigate the characteristics (photon occupation and statistics) of the coherent incident field which indirectly gives knowledge on the non-classical properties of the generated harmonics. Our approach is therefore important as it is applicable to a wide variety of centrosymmetric systems as the strong-coupling-induced symmetry breaking depends on the properties of photonic environment used to achieve strong light-matter interaction.

%\section{Theoretical framework}
%\label{sec:general-framework}

We start by describing the coupled light-matter system that considers transverse fields with wavelengths that are substantially larger than the matter system so that the long-wavelength limit~\cite{tannoudji1989,craig1998} is applicable. In this setting, the non-relativistic dynamics of the coupled system is described by the Pauli-Fierz Hamiltonian~\cite{rokaj2017,spohn2004}: 
\begin{align}
\hat{H} &= \sum\limits_{i=1}^{N}\left(\frac{\hat{\textbf{p}}_{i}^{2}}{2m} + v_{\textrm{ext}}(\hat{\textbf{r}}_i) - \frac{e}{m} \hat{\textbf{A}}\cdot\hat{\textbf{p}}_{i} + \frac{e^{2}}{2m} \hat{\textbf{A}}^{2}\right)  \nonumber \\
& \quad + \sum\limits_{i>j}^{N}w(\hat{\textbf{r}}_{i},\hat{\textbf{r}}_{j}) + \sum_{\alpha=1}^{M}\frac{1}{2}\left(\hat{p}_{\alpha}^{2} + \omega_{\alpha}^{2}\hat{q}^{2}\right) \, , \label{eq:velocity-gauge-hamiltonian}
\end{align}
where the $N$ electrons are described by the electronic coordinates, $\hat{\textbf{r}}_{i}$, the momentum operator of the electrons by $\hat{\textbf{p}}_{i}$, $w(\hat{\textbf{r}}_{i},\hat{\textbf{r}}_{j})$ describes the longitudinal interaction between the electrons and $v_{\textrm{ext}}(\hat{\textbf{r}}_i)$ represents the binding potential. The last term of Eq.~(\ref{eq:velocity-gauge-hamiltonian}) is the energy of the quantized electromagnetic field given in terms of the displacement coordinate $\hat{q}_{\alpha}$ and canonical momentum operator $\hat{p}_{\alpha}$ with associated mode frequency $\omega_{\alpha}$ for each mode $\alpha$ of an arbitrarily large but finite number of photon modes $M$. The vector potential is $\hat{\textbf{A}}=\sum_{\alpha=1}^{M}\boldsymbol{\lambda}_{\alpha}\hat{q}_{\alpha}$ where $\boldsymbol{\lambda}_{\alpha}=\sqrt{1/\epsilon_{0}V}\textbf{e}_{\alpha}$ is the coupling strength and $V$ is the quantization volume of the field. We note that in a standard investigation of harmonic generation based on a semi-classical description, the incident field is not treated as a quantized field but merely as an external time-dependent field that induces dynamics in the usual matter-only constituents~\cite{timurdogan2017,klein2017}. Here, we treat the incident pump field as a quantized field and investigate the changes the now self-consistently coupled light-matter system has on the symmetry and nonlinear properties of the system.

As an illustrative system with an inversion symmetry, we consider a well-studied semiconductor quantum ring of GaAs that features a single effective electron (see SI. Section~\ref{sup:quantum-ring} for further details). Due to the inversion symmetry of the system, nonlinear processes associated with even-order susceptibilities are forbidden~\cite{boyd1992}. However, we will demonstrate that the inversion symmetry of the system can be broken and controlled by coupling to a quantized photon mode which gives rise to efficient nonlinear optical processes like even-order harmonic generation. To achieve this, the matter system is placed inside an electromagnetic environment, for example, a high-Q optical cavity. The pump mode is treated as a quantized field that couples to the matter system as in equation (\ref{eq:velocity-gauge-hamiltonian}). We perform a numerically exact ab-initio simulation of the coupled light-matter system by solving the time-dependent Schr\"{o}dinger equation for the Hamiltonian of Eq.~(\ref{eq:velocity-gauge-hamiltonian}) (see SI. Section~\ref{sup:numerical-details} for details). As an initial state, we choose a factorizable product state of the ground-state of the matter system and the photon mode (pump field) initially prepared in a coherent state with $n=|\xi|^{2}=25$ photons where $\xi$ is the amplitude of the coherent state. The importance of a few-photon quantized pump field is that it presents an alternative way to achieve qualitatively the same result with small pump intensity (see SI. Section~\ref{sup:strong-field-vs-coupling} for details). We note that the choice of describing the pumping process with a coherent state yields quantitatively the same results as the case where the coupled system is driven by an external potential or charge current with the same amount of photons as demonstrated in Ref.~\cite{welakuh2021}.

\begin{figure}[bth] 
\centerline{\includegraphics[width=0.5\textwidth]{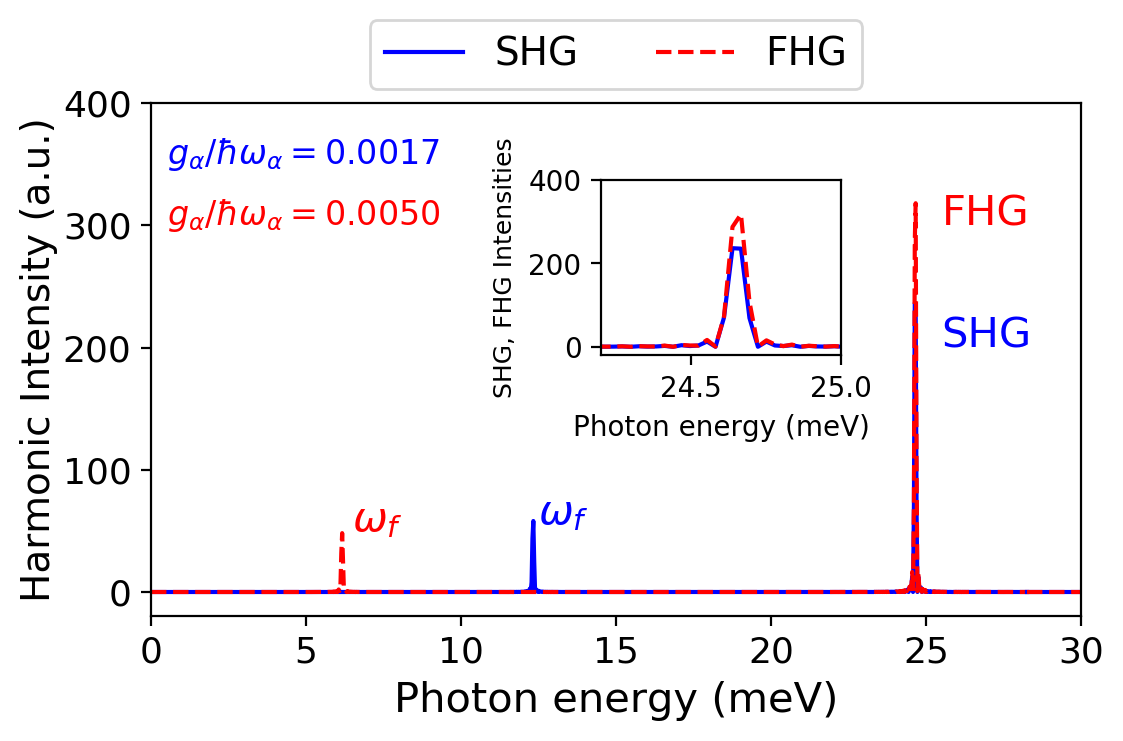}}
\caption{Induced SHG and FHG processes in an inversion symmetric system for a quantized treatment of the pump field in a coherent state. The inset shows that the harmonic intensity of the FHG is larger than the SHG process which is attributed to the strong light-matter coupling. $\omega_{f}$ are the fundamental (pump) frequencies for SHG and FHG.}
\label{fig:GaAs-even-harmonic-spectrum}
\end{figure}

For the coupled light-matter system, we now investigate individually the nonlinear optical processes of second-harmonic generation (SHG) and fourth-harmonic generation (FHG) that are normally forbidden in such a system. To realize these nonlinear processes in a quantized light-matter system, the fundamental pump frequency is chosen to be one-half (SHG) and one-fourth (FHG) the transition energy (24.65~meV) between the electronic states $|\varphi_{1}^{0}\rangle\leftrightarrow|\varphi_{7}^{1}\rangle$ (see SI. Section~\ref{sup:quantum-ring}). The results of this calculation is shown in Fig.~(\ref{fig:GaAs-even-harmonic-spectrum}). As the quantized pump mode breaks the centrosymmetry of the matter system due to strong light-matter coupling (see SI. Section~\ref{sup:quantum-ring} for details), this allow for forbidden nonlinear processes such as SHG and FHG to now exist in the coupled system. Thus, this highlights a strong-coupling-induced symmetry breaking in centrosymmetric systems. In the standard semi-classical description of harmonic generation, for a fixed pump field the intensity of the generated harmonics (second-, third-, fourth-harmonics, etc) decrease as a result of the associated nonlinear susceptibilities~\cite{boyd1992}. Interestingly, we find in the fully quantized treatment that for a fixed amplitude of the coherent state $\xi=5$, the intensity of the FHG is larger than the SHG process. This increase is attributed to the light-matter coupling strength 
\begin{align}
g_{\alpha,ij}^{(kl)} = \frac{e}{m}\sqrt{\frac{\hbar}{2\epsilon_{0}\omega_{\alpha}V}} \langle \varphi_{i}^{k} | \textbf{e}_{\alpha}\cdot\hat{\textbf{p}} |\varphi_{j}^{l}\rangle \, , \label{eq:light-matter-coupling}
\end{align}
where $|\varphi_{j}^{l}\rangle$ are the eigenstates of the matter system and $\omega_{\alpha}$ in this case represents the fundamental frequency of the SHG and FHG processes. Typically, in a strong light-matter coupling regime there is an emergence of polariton states due to a resonant coupling of a photon mode with an electronic resonance. In our investigation the contribution from polariton states do not play a significant role since the transition amplitudes involving these states are negligible as the fundamental frequency is off-resonance for the respective harmonic generation processes. Clearly, for a fixed volume $V$ the coupling $g_{\alpha,ij}^{(kl)}$ (and vector potential) is greater for FHG than it is for the SHG process. The difference in the light-matter coupling for these different processes accounts for the FHG having an intensity larger than the SHG as shown in Fig.~(\ref{fig:GaAs-even-harmonic-spectrum}). It is important to note that we cannot just increase the coupling strength arbitrary by reducing the frequency to zero since there is a natural renormalization in the lower frequency that arises as an intrinsic frequency cutoff~\cite{rokaj2020,spohn2004}. This intrinsic cutoff which has a dependence on the number of particles ensures that no infrared divergence appears. However, unphysical divergences may arise if this natural renormalization of the frequencies is not taken properly into account. 

\begin{figure}[bth] 
\centerline{\includegraphics[width=0.5\textwidth]{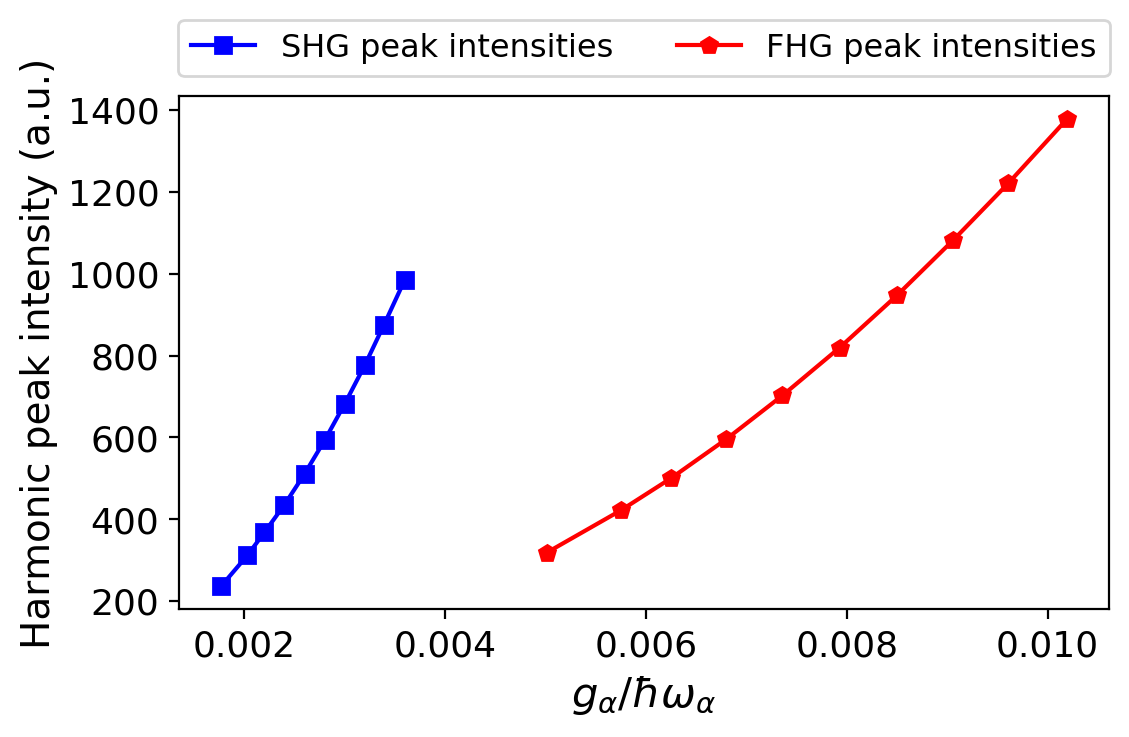}}
\caption{Intensity dependence of even-order harmonic generation where the harmonic intensity increases for increasing light-matter coupling. Here $g_{\alpha}$ increases by decreasing $V$ while $\omega_{\alpha}$ have values for the pump frequencies for SHG and FHG. The intensity of FHG is larger when compared to SHG due to stronger coupling $g_{\alpha}/\hbar\omega_{\alpha}$.}
\label{fig:GaAs-coupling-spectrum}
\end{figure}

Another means by which the light-matter coupling can be increased for the single emitter case here is by reducing the volume $V$ as normally done in other theoretical works~\cite{flick2019,welakuh2022,welakuh2022a} and realized in experiments~\cite{barachati2018}. The  controlled strong-coupling-induced symmetry breaking in the coupled system is linked to  changing the coupling strength by reducing the cavity volume (see App.~\ref{sup:quantum-ring}). As the interaction between the photon field and matter breaks the inversion symmetry for a chosen amplitude of the incident field, a new possibility arises to control the degree of the induced asymmetry by strongly coupling the system. This is accompanied by a significant increase of the intensity of the induced harmonic generation process. To illustrate this, we show the peak intensity dependence of the harmonic generation processes on the dimensionless ratio $g_{\alpha,ij}^{(kl)}/\hbar\omega_{\alpha}$ which provides an indication for relative light-matter coupling strengths. In Fig.~(\ref{fig:GaAs-coupling-spectrum}), for the same decreasing values of $V$ in Eq.~(\ref{eq:light-matter-coupling}) (and increasing $g_{\alpha,ij}^{(kl)}/\hbar\omega_{\alpha}$) for both processes, we find that the peak intensities increase significantly for both SHG and FHG processes. This result shows clearly how the SHG and FHG processes can be made more efficient by increasing the light-matter coupling strength. An added merit of the quantized treatment is that we can investigate the characteristics (photon occupation and statistics) of the coherent pump field which indirectly provides information on the non-classical properties of the generated harmonics as discussed in App.~\ref{sup:photon-field}. It is important to note that in the decoupling limit of the system, i.e., in the limit where $g_{\alpha}$ tends to zero (free space coupling) for very large volumes, the inversion symmetry can no longer be broken and the induced nonlinear optical processes do not occur.

\begin{figure}[bth] 
\centerline{\includegraphics[width=0.5\textwidth]{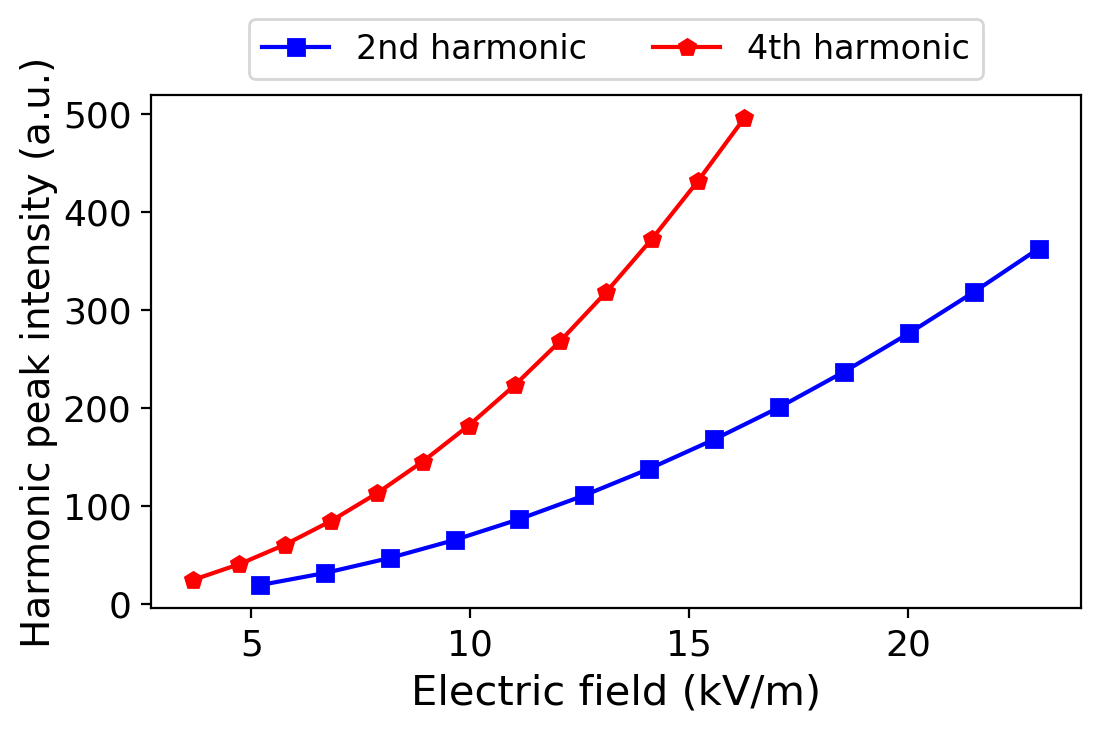}}
\caption{The pump field dependence on the intensity of the induced even-order harmonic generation processes in an inversion symmetric system for a quantized treatment of the coupled system. The electric field has a quadratic (quartic) dependence on the peak intensities of SHG (FHG).}
\label{fig:GaAs-power-spectrum}
\end{figure}

\begin{figure}[bth] 
\centerline{\includegraphics[width=0.5\textwidth]{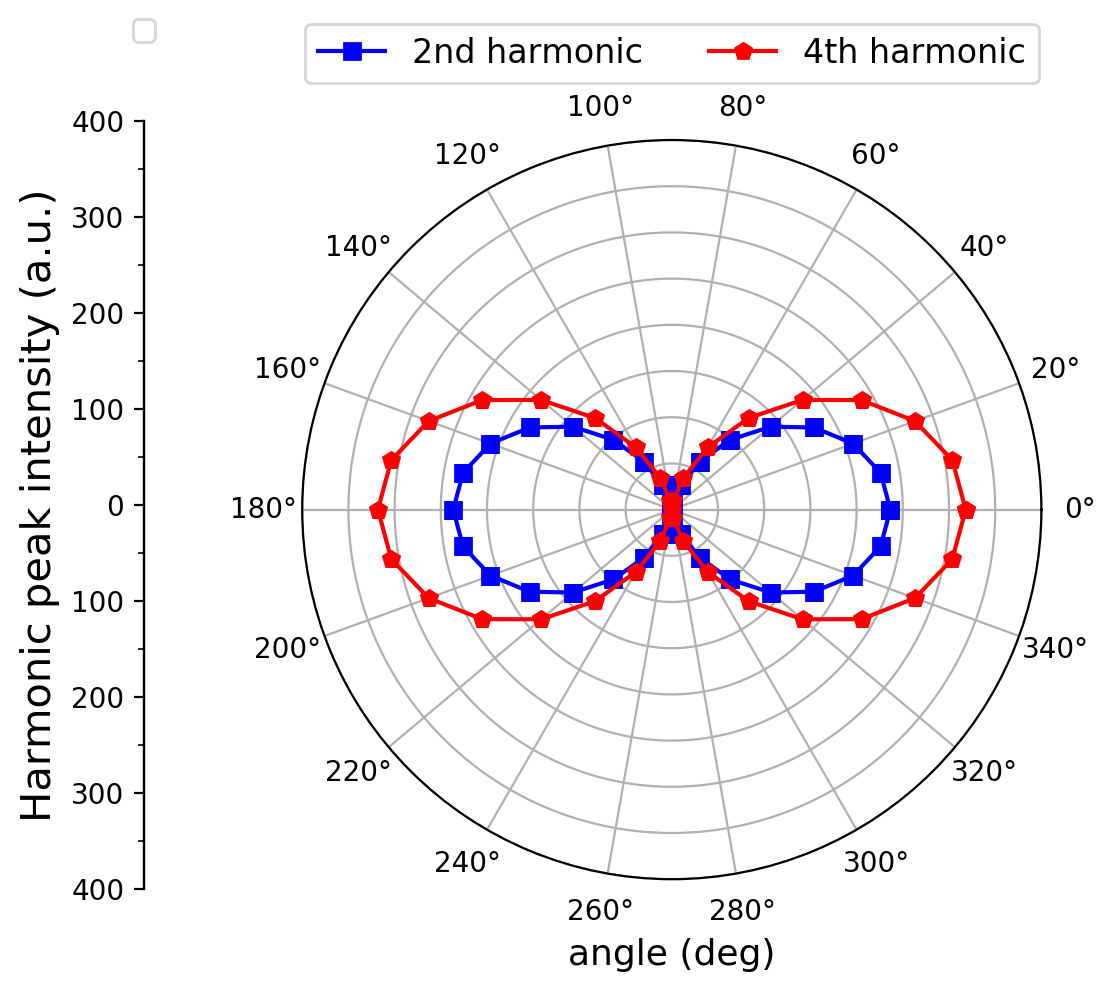}}
\caption{Polar plot of the polarization dependence of the induced even-order harmonic generation intensities. The radiation pattern for different polarization angles $\theta$ shows the intensities varies as $I_{\text{SHG,FHG}} \propto \cos^{2}(\theta)$. The polar angle $\theta$ is defined with respect to the $x$-axis of the quantum ring.}
\label{fig:GaAs-polar-spectrum}
\end{figure}

In the next step, we verify that the even-order harmonic processes in Fig.~(\ref{fig:GaAs-even-harmonic-spectrum}) are truly the SHG and FHG processes. To do this, we calculate the dependence of their peak intensities on the incident field by varying the pump amplitude $\xi$ from $1.4$ to $6.2$ with an equidistant spacing $\Delta \xi = 0.4$ for a fixed cavity volume. In Fig.~(\ref{fig:GaAs-power-spectrum}), we show the peak intensity dependence of the SHG (FHG) on the electric field of the pump mode which has a quadratic (quartic) dependence, confirming that these processes are indeed the SHG and FHG. Of particular interest is the dipole radiation pattern of a system whose inversion symmetry has been broken (reduced), in this case, due to coupling to a quantized pump field. For this, we calculate the polarization dependence spectra of the even-order harmonic generation in Fig.~(\ref{fig:GaAs-even-harmonic-spectrum}). The polarization of the incident field which was originally chosen such that the angle $0^{\circ}$ corresponds to the polarization direction along the $x$-axis is now varied. In Fig.~(\ref{fig:GaAs-polar-spectrum}) we show a polar plot of the calculated peak intensity dependence on the pump polarization where we find that the induced SHG and FHG is maximum when the radiated field is parallel to the pump field. Also, as the pump and harmonic processes are collinearly polarized, we obtain a radiation pattern where the intensities varies as $I_{\text{SHG,FHG}} \propto \cos^{2}(\theta)$ where $\theta$ is the incident polarization angle. This is expected for a linearly polarized pump field along a single direction (axis).

\begin{figure}[bth] 
\centerline{\includegraphics[width=0.5\textwidth]{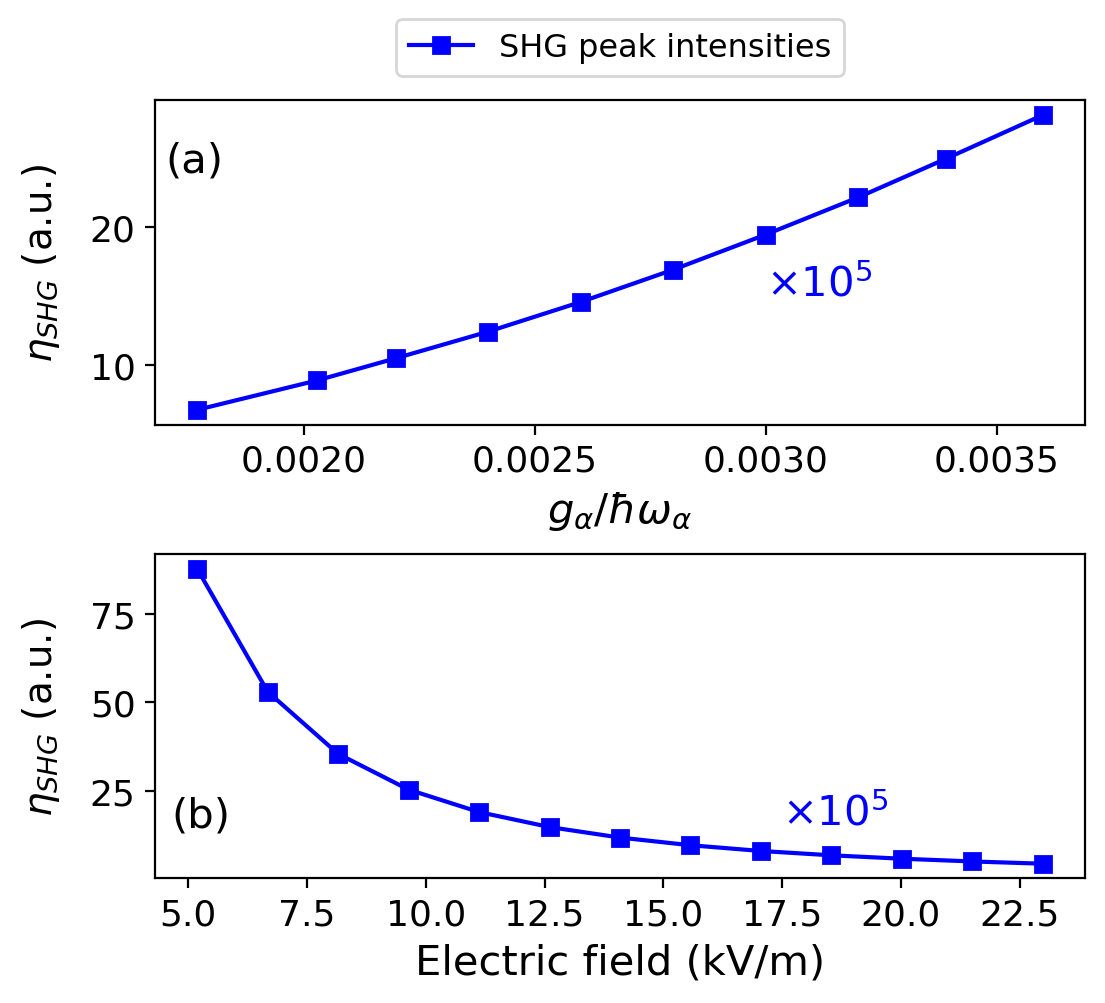}}
\caption{(a) Induced SHG efficiency dependence on the coupling strength which shows an increase for increasing $g_{\alpha}$. (b) A decrease in the conversion efficiency of SHG for increasing pump field strengths which shows the efficiency will eventually saturate. A comparison between (a) and (b) indicates strong coupling is favorable over using strong fields.}
\label{fig:GaAs-efficiency-spectrum}
\end{figure}

We now consider an important aspect of the harmonic generation which is the nonlinear conversion efficiency. It is clear that the SHG and FHG processes are efficient from a centrosymmetric system with broken symmetry as opposed to the case where the symmetry remains intact since $\chi^{(n)}=0$ (for even $n$). Beyond this observation, we show that our approach (which is the few-photon limit and strong coupling) is favorable over the standard approach which requires strong fields. Towards this, we consider the case of SHG where the nonlinear conversion efficiency is defined as $\eta_{\text{SHG}} = I_{2\omega}/I_{\omega}^{2}$ where $I_{\omega}$ is the peak intensity of the fundamental frequency. To compare between the two methods, we choose the pump amplitude $\xi=5$ (which gives an electric field strength $E_{0}=18.54$~kV/m) and vary the coupling strength $g_{\alpha}/\hbar\omega_{\alpha}$ for the strong coupling case and for the strong field case, we fix the coupling $g_{\alpha}/\hbar\omega_{\alpha}=0.0017$ and vary the electric field strength by increasing $\xi$. The results of this comparison are shown in Fig.~(\ref{fig:GaAs-efficiency-spectrum}) where we find the efficiency increases for the strong coupling case but decreases for increasing electric field strengths of the pump. This result highlights that we do not necessarily need an intense pump field for the harmonic generation process but rather need strong light-matter coupling as also demonstrated for the case of the down-conversion process~\cite{welakuh2021}. It equally validates the long-standing goal in optical science to implement nonlinear optical effects (here the harmonic generation) at progressively lower pump powers or pulse energies~\cite{chang2014}.

%\section{Conclusion and outlook}
%\label{sec:conclusion-outlook}

To conclude, in this work we have demonstrated a general approach to break the inversion symmetry of material systems, thus enabling induced and highly tunable nonlinear optical processes such as SHG and FHG, typically physically forbidden in centrosymmetric systems. Our approach relies on a quantum mechanical treatment of the coupled light-matter system where the pump field is treated as a quantized field in a coherent state. We demonstrate that the inversion symmetry can be broken by strongly coupling the light-matter, thus allowing for even-order harmonic generation processes to occur. In particular, we show how the strong-coupling-induced symmetry breaking can be controlled by increasing the light-matter coupling strength which simultaneously results to efficient SHG and FHG. Such control is of great interest for several applications ranging from optical frequency conversion to imaging and bio-sensing where high efficiency is required~\cite{garmire2013}. Our choice of a matter system with a single effective electron interacting with the photon field was to illustrate in more detail the strong-coupling-induced symmetry breaking that results to even-order processes like SHG and FHG in centrosymmetric systems. In the case where we have an ensemble of emitters or a general many-electron system, the light-matter coupling strength increases by a square-root of the number of identical emitters~\cite{garraway2011}. In such multi-electron systems, the efficiency and induced harmonic yield of the SHG and FHG will increase significantly due to collective strong coupling. Such many-electron systems interacting with the quantized electromagnetic field can be treated with already existing first-principles methods for strongly coupled light-matter systems~\cite{flick2018,svendsen2021,haugland2020,jestaedt2020}. Our approach is applicable to a wide variety of material systems since the strong-coupling-induced symmetry breaking relies primarily on the ability to tune properties of the electromagnetic environment to attain strong light-matter interaction. Further, this approach is not limited to optical cavities but applicable in other photonic environments used to achieve strong light-matter coupling. We note that strong-coupling induced symmetry breaking in centrosymmetric systems allows for other nonlinear optical effects associated with even-order susceptibilities other than the harmonic generation process. Finally, this approach can be applied to reduce other symmetries of  material systems to control physical and chemical processes.

%\section{Acknowledgments}

\emph{Acknowledgments:} We acknowledge helpful discussions with Nicolas Tancogne-Dejean, Jonathan Curtis, Stephane Kena-Cohen, and Aaron Lindenberg. This work is primarily supported through the Department of Energy BES QIS program on `Van der Waals Reprogrammable Quantum Simulator' under award number DE-SC0022277 for the work on nonlinearity in low dimensional systems, as well as partially supported by the Quantum Science Center (QSC), a National Quantum Information Science Research Center of the U.S. Department of Energy (DOE) on cavity-control of quantum matter. P.N. acknowledges support as a Moore Inventor Fellow through Grant No. GBMF8048 and gratefully acknowledges support from the Gordon and Betty Moore Foundation as well as support from a NSF CAREER Award under Grant No. NSF-ECCS-1944085.
\vspace{-2em}

% ----------------------------------------------------------------
%  Bibliography                            
% ----------------------------------------------------------------

\vspace{10em}

\bibliography{01_light_matter_coupling} % Produces the bibliography via BibTeX.

%\appendix

% ----------------------------------------------------------------
%  Merge with supplemental materials                          
% ----------------------------------------------------------------

\clearpage
\pagebreak

\onecolumngrid
\widetext

\begin{center}
\textbf{\large Supplemental Information:\\{Nonlinear optical processes in centrosymmetric systems by  strong-coupling-induced symmetry breaking
}} \\
[1em] 
{Davis M. Welakuh and Prineha Narang} \\
\textit{Harvard John A. Paulson School Of Engineering And Applied Sciences, Harvard University, Cambridge, Massachusetts 02138, USA}
\end{center}

% ----------------------------------------------------------------
% Prefix a "S" to all equations, figures, tables and reset the  counter                         
% ----------------------------------------------------------------

\setcounter{equation}{0}
\setcounter{figure}{0}
\setcounter{table}{0}
\setcounter{page}{1}
\makeatletter
\renewcommand{\theequation}{S\arabic{equation}}
\renewcommand{\thefigure}{S\arabic{figure}}
% \renewcommand{\bibnumfmt}[1]{[S#1]}
% \renewcommand{\citenumfont}[1]{S#1}

% Start two columns
\twocolumngrid

\section{2D semiconductor quantum ring}
~\label{sup:quantum-ring}

\begin{figure}[bth]
\centerline{\includegraphics[width=0.5\textwidth]{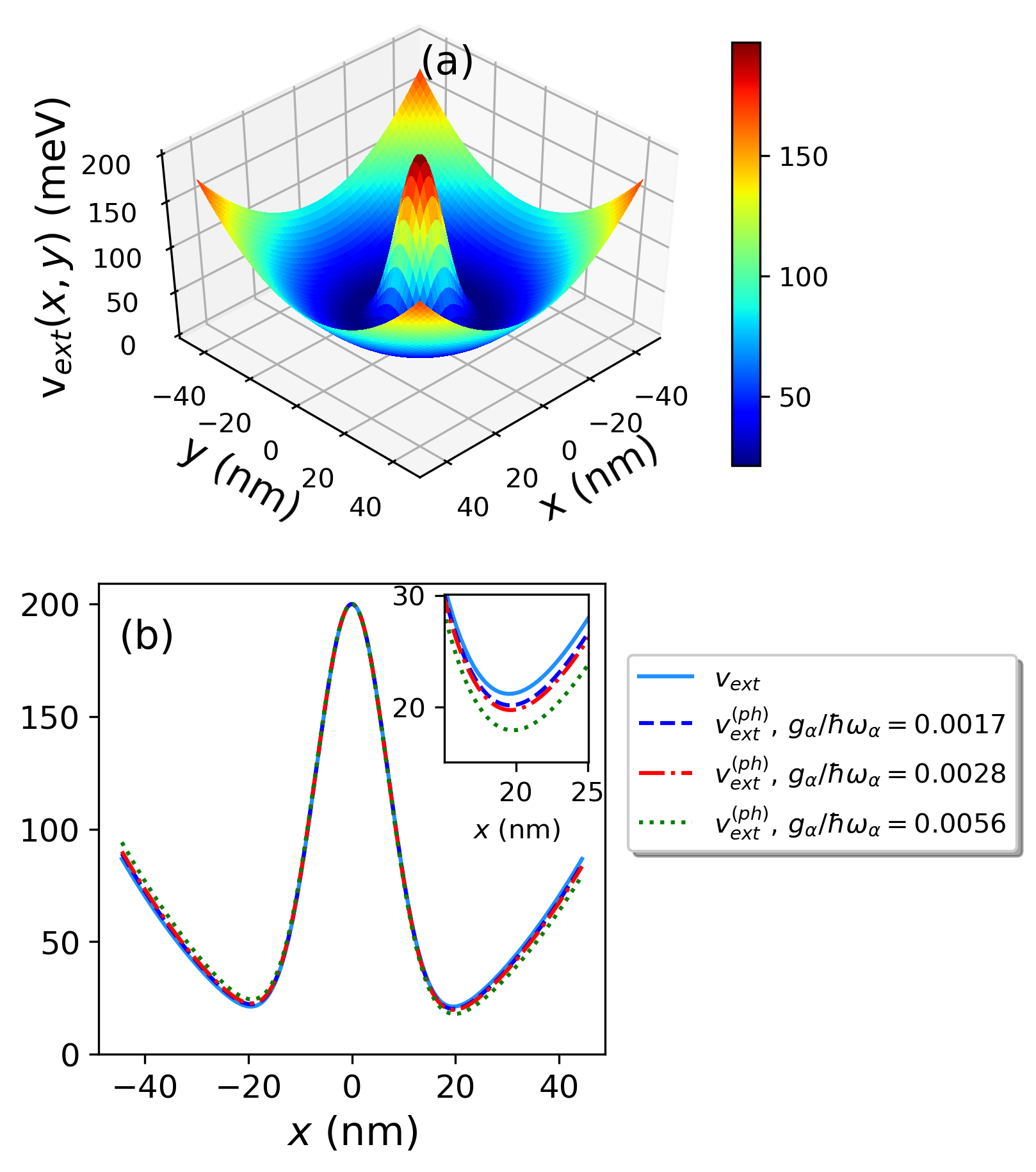}}
\caption{(a) Real-space 2D potential of the quantum ring with a Gaussian peak at the center. The potential is symmetric with respect to its center of inversion. (b) Comparison of the bare quantum ring potential to the case when coupled to a quantized field along the $x$-direction for decreasing cavity volumes (increasing coupling strength).}
\label{fig:potential}
\end{figure}

\begin{figure}[bth]
\centerline{\includegraphics[width=0.5\textwidth]{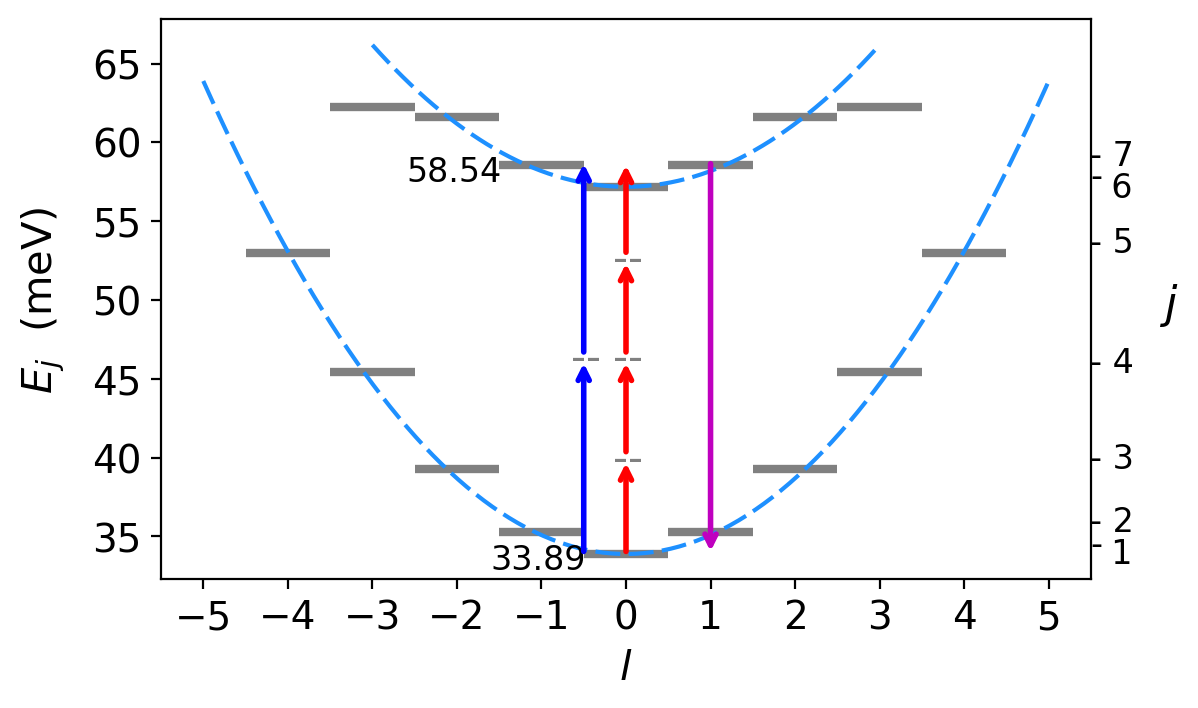}}
\caption{Energy spectrum of the 2D quantum ring showing the two lowest radial bands with degenerate and nondegenerate excited states and a nondegenerate ground state for different angular momentum $l$. The blue and red arrows indicate the SHG and FHG which yields a harmonic field (purple) which is a multiple of their respective pump energies.}
\label{fig:GaAs-spectrum-harmonics}
\end{figure}

The matter system is a 2D semiconductor quantum ring of finite width that features a single effective electron confined in two-dimensions in real-space ($\hat{\textbf{r}} = \hat{x} \textbf{e}_{x} + \hat{y} \textbf{e}_{y}$). The system is described by the Hamiltonian
\begin{equation}
\hat{H}_{\text{M}} = -\frac{\hbar^{2}}{2m} \left(\frac{\partial^{2}}{\partial x^{2}} + \frac{\partial^{2}}{\partial y^{2}}\right)  + \underbrace{\frac{1}{2}m\omega_{0}^{2}\hat{\textbf{r}}^{2} + V_{0} e^{-\hat{\textbf{r}}^{2}/d^{2}}}_{v_{\text{ext}}(\textbf{r})} , \label{eq:electron-hamiltonian}
\end{equation}
where the potential part with $\hat{\textbf{r}}^{2} = \hat{x}^{2} + \hat{y}^{2}$ introduces a parabolic confinement including a Gaussian peak located at the center as shown in Fig.~(\ref{fig:potential} a). The parameters of the potential are chosen such that it reflects the energy and length scales of a semiconductor quantum ring of GaAs as used in experiments~\cite{fuhrer2001,ihn2005} where $\hbar \omega_{0} = 10$ meV, $d = 10$ nm, $m = 0.067 m_{e}$, and $V_{0} =200$ meV. In this regard, we work in scaled effective atomic units: $\text{Ha}^{*}=(m/\epsilon^{2})\text{Ha} \approx 11.30$~meV, $a_{B}^{*}=(m/\epsilon)a_{0}\approx10.03$~nm, and $u_{t}^{*}=\hbar/\text{Ha}^{*}\approx 58.23$~fs where we made use of the dielectric constant $\epsilon=12.7\epsilon_{0}$ of GaAs~\cite{rasanen2007}. The eigenstates $|\varphi_{j}^{l}\rangle$ are labeled by the angular momentum $l=0,\pm 1, \pm 2, \pm 3, ... \,$ and the index $j=|l|+1$ enumerates over the energy levels (see Fig.~(\ref{fig:GaAs-spectrum-harmonics})). Due to symmetry of the system, dipole-allowed transitions can only occur between states with consecutive angular momenta~\cite{rasanen2007,flick2015,welakuh2021}.

To understand how the photon mode breaks the symmetry, we consider the electron-photon system in the unitary equivalent length gauge~\cite{rokaj2017}. For the quantum ring coupled to one mode, the Hamiltonian has the form:
\begin{align}
\hat{H}_{\text{L}} &= -\frac{\hbar^{2}}{2m} \left(\frac{\partial^{2}}{\partial x^{2}} + \frac{\partial^{2}}{\partial y^{2}}\right) + \frac{1}{2} \left(\hat{p}^{2} + \omega^{2}\hat{q}^{2}\right)  \nonumber \\
& \quad + \underbrace{\frac{1}{2}m\omega_{0}^{2}\hat{\textbf{r}}^{2} + V_{0} e^{-\hat{\textbf{r}}^{2}/d^{2}} - e \, \omega \, \hat{q} \, \boldsymbol{\lambda}\cdot\hat{\textbf{r}} + \frac{1}{2}\left(e \, \boldsymbol{\lambda}\cdot\hat{\textbf{r}}\right)^{2} }_{v_{\text{ext}}^{(ph)}(\textbf{r})} \, . \label{eq:length-gauge-hamiltonian}
\end{align}
The dipole self-energy (last term) adds to the harmonic binding potential (i.e. $\frac{1}{2}m\omega_{0}^{2}\hat{\textbf{r}}^{2}$) and the bilinear interaction term (i.e. $e \,\omega \, \hat{q} \, \boldsymbol{\lambda}\cdot\hat{\textbf{r}}$) breaks the symmetry of the quantum ring for a specific polarization of the cavity mode. This is clear since $v_{\text{ext}}^{(ph)}(-\textbf{r})\neq v_{\text{ext}}^{(ph)}(\textbf{r})$. Besides breaking the symmetry, there is the possibility to control the degree of asymmetry by reducing the cavity volume. In  Fig.~(\ref{fig:potential} b), we show how to control the asymmetry by changing the cavity volume $V$.

\section{Light-matter coupling strength and the quantized pump field}
\label{sup:photon-field}

Provided we can obtain all electronic eigenenergies $E_{i}$ and eigenstates $|\psi_{i}\rangle$ of a matter system, for example through numerical exact diagonalization, we can write velocity gauge Hamiltonian Eq.~\ref{eq:velocity-gauge-hamiltonian} as:
\begin{align} 
\hat{H} &=\sum\limits_{i=1}\hbar\omega_{i} |\psi_{i}\rangle\langle \psi_{i} | +\sum_{\alpha=1}^{M}\frac{1}{2}\left(\hat{p}_{\alpha}^{2}+\omega_{\alpha}^{2}\hat{q}_{\alpha}^{2} \right) \nonumber\\ 
&\quad - \sum_{i,j=1}\sum_{\alpha=1}^{M} g_{\alpha}^{(ij)} \left(\hat{a}_{\alpha} + a_{\alpha}^{\dagger}\right) |\psi_{i}\rangle\langle \psi_{j} | + \frac{e^{2}}{2m}\hat{\textbf{A}}^{2}
\, . \label{eq:el-pt-hamiltonian-energy-basis}
\end{align}
The the electron-photon coupling term in the above equation is defined to be
\begin{align}
g_{\alpha}^{(ij)} = \frac{e}{m}\sqrt{\frac{\hbar}{2\epsilon_{0}\omega_{\alpha}V}} \langle \psi_{i} | \textbf{e}_{\alpha}\cdot\hat{\textbf{p}} |\psi_{j}\rangle \, ,
\end{align}
Here, we consider a linearly polarized field with polarization $\textbf{e}_{\alpha} = \cos\theta\textbf{e}_{x} + \sin\theta\textbf{e}_{y}$. From the operator of the quantized vector potential
\begin{align}
\hat{\textbf{A}} &=\sum_{\alpha=1}^{M}\sqrt{\frac{\hbar}{2\epsilon_{0}\omega_{\alpha}V}}\left(\hat{a}_{\alpha} + \hat{a}_{\alpha}^{\dagger}\right) \, \textbf{e}_{\alpha} \, ,
\end{align}
we obtain the vector potential at the initial time by taking the expectation with the coherent state:
\begin{align}
\langle\xi|\hat{\textbf{A}}|\xi\rangle &=\sum_{\alpha=1}^{M}\sqrt{\frac{\hbar}{2\epsilon_{0}\omega_{\alpha}V}} \, 2|\xi|\cos\phi \, \textbf{e}_{\alpha} \, .
\end{align}
The coherent state has the form $|\xi\rangle= e^{-|\xi|^{2}/2} \sum_{n=0}^{\infty}  \left(\xi^{n}/\sqrt{n!}\right)|n\rangle$ where $\xi = |\xi|e^{i\phi}$ with $|\xi|$ being the amplitude, $\phi$ is the phase and $|n\rangle$ the Fock number state of the photon mode. At the initial time, there are $n=\langle\xi|\hat{a}^{\dagger}\hat{a}|\xi\rangle=|\xi|^{2}$ number of photons in the pump field.

\begin{figure}[bth]
\centerline{\includegraphics[width=0.5\textwidth]{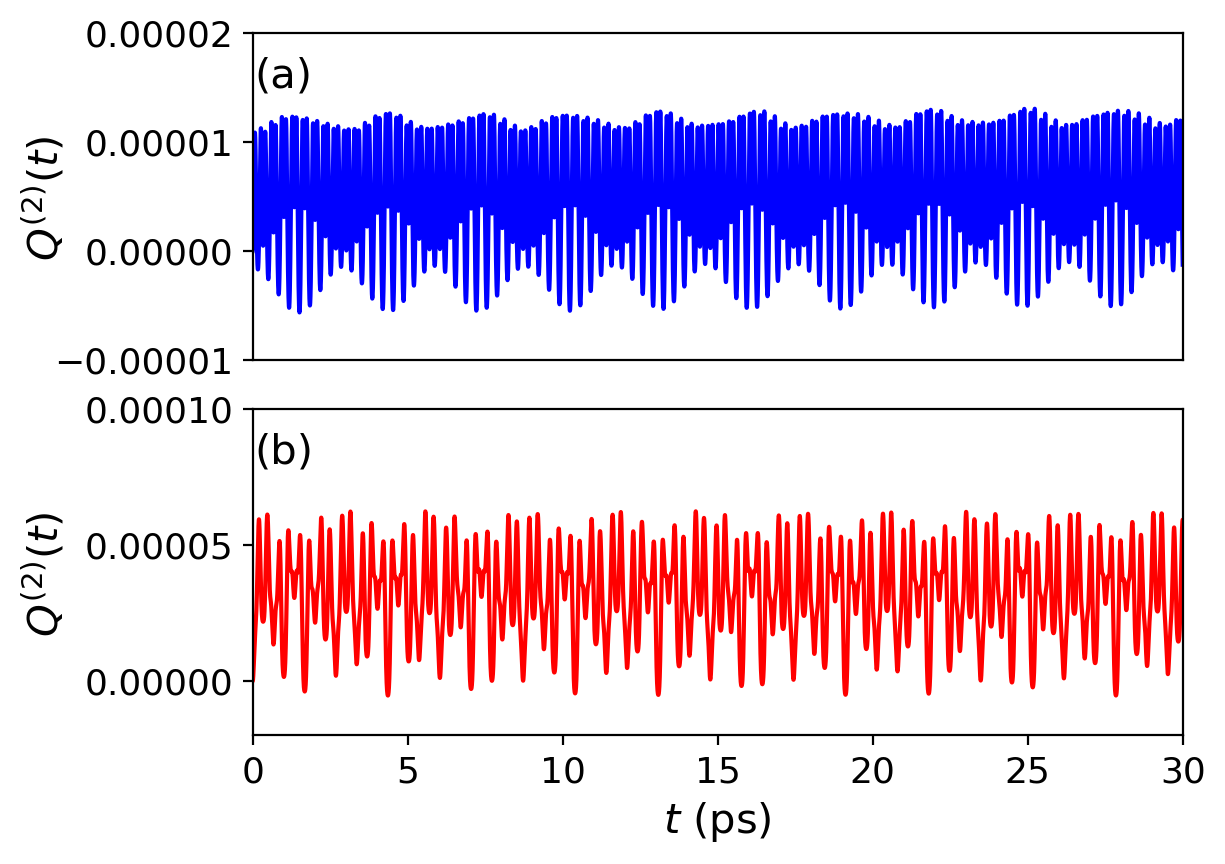}}
\caption{The time-evolved Mandel $Q$ parameter for (a) SHG and (b) FHG for an amplitude $\xi=2$. The photon statistics of SHG varies between a bunching and anti-bunching character. FHG shows mainly super-Poissonian statistics.}
\label{fig:GaAs-mandel-Q-parameter}
\end{figure}

Since the pump field is treated as a quantized field, we can investigate its photon statistics. We determined this property by computing the Mandel $Q_{\alpha}$ parameter~\cite{mandel1979} defined as
\begin{align}
Q_{\alpha} =  \frac{\langle \hat{a}_{\alpha}^{\dagger}\hat{a}_{\alpha}^{\dagger}\hat{a}_{\alpha}\hat{a}_{\alpha}\rangle - \langle \hat{a}_{\alpha}^{\dagger}\hat{a}_{\alpha}\rangle^{2} }{\langle \hat{a}_{\alpha}^{\dagger}\hat{a}_{\alpha}\rangle} \, ,  \label{mandel-q}
\end{align}
where $\hat{a}_{\alpha}^{\dagger}$ and $\hat{a}_{\alpha}$ are the creation and annilation operators of the $\alpha$ mode. The parameter $Q_{\alpha}$ measures the deviation of the photon statistics from a Poisson distribution and thus is a measure for the non-classicality. A field with non-classical properties have a range of values between $-1 \leq Q_{\alpha}< 0$ which corresponds to sub-Poissonian statistics (anti-bunching behavior). Fields with super-Poissonian statistics (bunching behavior) have $Q_{\alpha} > 0$ and for a coherent state with Poissonian statistics we have $Q_{\alpha} = 0$ \cite{loudon2000}. Figure (\ref{fig:GaAs-mandel-Q-parameter}) shows super-Poissonian statistics for FHG since it couples weakly and for SHG, it shows statistics that vary between a bunched and antibunhed field. This is due to the coherent exchange of energy between the pump mode and the matter system. From the non-classical characteristic of the coherent pump field, we infer that the generated second-harmonic will have similar features as the coupled system coherently interacts and exchange energy. Our strongly coupled light-matter treatment brings to light new observables that are not usually considered in SHG or FHG since the pump field is treated classically. Furthermore, it highlights the possibility where the generated harmonics have non-classical signatures which is important for applications in quantum-information-processing protocols~\cite{obrien2009}.

\section{Strong fields versus strong coupling}
\label{sup:strong-field-vs-coupling}

In this section, we investigate the outcome of the harmonic generation process when the system is excited with a strong intense field versus a few-photon field in the strong coupling regime. To do this, we perform two simulations where the case involving the few-photon field is treated as describe above and the case involving a strong intense field is treated with a semi-classical description. This means that in Eq.~(\ref{eq:velocity-gauge-hamiltonian}) the contribution of the electromagnetic energy (last term) is dropped out and the vector potential is simply a time-dependent external field that perturbs the system which has the following form
\begin{align}
\textbf{A}(t) &= \frac{\text{E}_{0}}{\omega}\cos(\omega t)\sin^{2}\left(\pi\frac{t}{\tau}\right) \textbf{e}_{x} \, , \quad 0 \leq t \leq \tau \, , \label{eq:electric-field}
\end{align}
where $\omega$ is the fundamental frequency, $E_{0}$ is the amplitude. As we work in effective atomic units, it is important to present the rescaled quantities. The electric field is $E^{*}=E_{h}^{*}/e \, a_{0}^{*}=((0.067)^{2}/(12.7)^{3})(E_{h}/e \, a_{0})$ which gives an intensity $I^{*}=1/2 \, c \, \epsilon E^{*2}=((0.067)^{4}/(12.7)^{5})\, I$ where $(E_{h}/e \, a_{0})=5.1422\times 10^{11}$~V/m and $I=3.51\times 10^{16}$~W/cm$^{2}$. For a strong field, we use an intensity $I_{0}=2.74\times 10^{4}$~W/cm$^{2}$ ($E_{0}=1.27\times 10^{5}$~V/m) which corresponds to a typical intensity needed for pronounced dipole oscillations for quantum rings~\cite{fomin2018}. For the few-photon case for strong coupling, the electric field is $E_{0}=7.41\times 10^{3}$~V/m where $\xi = 2$.

\begin{figure}[bth]
\centerline{\includegraphics[width=0.5\textwidth]{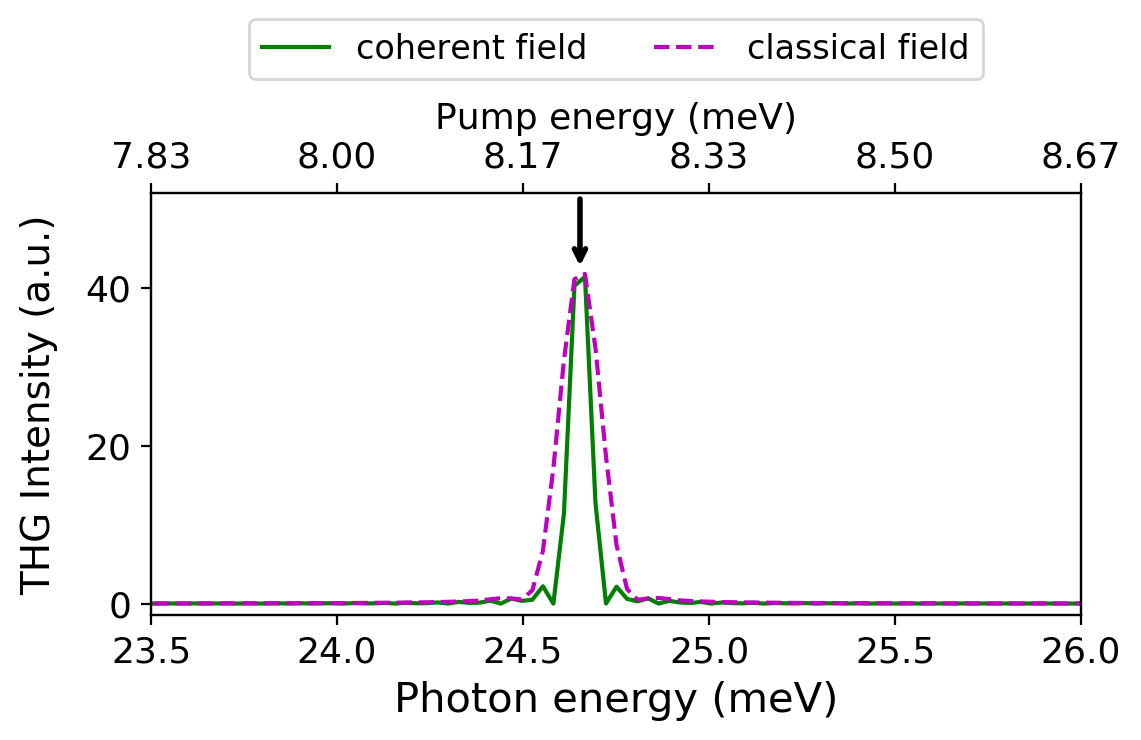}}
\caption{Strong field versus strong coupling for the case of THG which yields the same harmonic intensity. The strong field case has an electric field $E_{0}=1.27\times 10^{5}$~V/m while the strong coupling (with $g_{\alpha}/\hbar\omega_{\alpha}=0.0017$) case has an electric field $E_{0}=7.41\times 10^{3}$~V/m where $\xi = 2$. }
\label{fig:GaAs-THG-SF-vs-SC}
\end{figure}

To make a concrete comparison for both cases, we investigate the third-harmonic generation (THG) which does not depend on the symmetry of the material system. This at the same time shows that our method is applicable for noncentrosymmetric systems where THG is present. To simulate this nonlinear optical process, the pump energy is one-third the transition energy (24.65 meV). In Fig.~(\ref{fig:GaAs-THG-SF-vs-SC}), we show the strong field versus strong coupling for the case of THG. By comparing the two approaches, we find that for the strong field case with electric field $E_{0}=1.27\times 10^{5}$~V/m and strong coupling (with $g_{\alpha}/\hbar\omega_{\alpha}=0.0017$) case with electric field $E_{0}=7.41\times 10^{3}$~V/m where $\xi = 2$,  we obtain the same harmonic intensity for both approaches. Yet, the electric field (and intensity) for the strong field case is larger than that for the strong coupling case with just 4 photons in the pump field at the initial time. This highlights that strong coupling in the few-photon limit is a suitable approach to achieve an efficient nonlinear optical process.

\section{Numerical Details}
\label{sup:numerical-details}

The numerical details to treat the matter-photon coupled system is outline here. The matter Hamiltonian of Eq.~(\ref{eq:electron-hamiltonian}) is represented on a two-dimensional uniform real-space grid of $N_{x} = N_{y} = 127$ grid points (implying $127^{2}$ states are considered) with spacing $\Delta x = \Delta y = 0.7052$ nm while applying an eighth-order finite-difference scheme for the momentum operator and Laplacian. The photon mode that represent the pump field is represented in a basis of Fock number states. To represent the coherent state with $n(t_{0})=|5|^2$ photons, we include 65 Fock number states. The combined electron-photon space has dimensions $127\times 127 \times 65$ and we explicitly construct matrix representations for all operators in the Hamiltonian and observables of interest. To solve the time-dependent Schr\"{o}dinger equation of the coupled system, we use a Lanczos propagation scheme~\cite{hochbruck1997}. We compute expectation values of the observables for every time step $\Delta t = 0.0012$ ps of the time-evolved wave function.

The harmonic spectrum of the coupled system is calculated from the accelerated dipole response. For a laser-driven system, the experimentally measured harmonic spectrum is proportional to the modulus squared of the Fourier transformed laser-induced dipole acceleration~\cite{schafer1997,sundaram1990}:
\begin{align}
H(\omega) = \left| \int_{0}^{T} dt \frac{d^{2}}{dt^{2}} \textbf{r}(t) \, e^{-i\omega t} \right|^{2} . \label{eq:harmonic-spectrum}
\end{align}

\end{document}